\documentclass[11pt,a4paper]{article}

\usepackage[utf8]{inputenc}
\usepackage[T1]{fontenc}
\usepackage{lmodern}
\usepackage{microtype}

\usepackage{amsmath,amssymb,amsthm}

\usepackage{graphicx}
\usepackage{algorithm}
\usepackage{algpseudocode}

\usepackage[a4paper,margin=1in]{geometry}
\usepackage{xspace}
\usepackage{comment}

\usepackage[authoryear,round]{natbib}
\usepackage{hyperref}

\theoremstyle{plain}

\newtheorem{theorem}{Theorem}[section]

\theoremstyle{definition}

\DeclareMathOperator*{\argmax}{arg\,max}

\title{
Highly Scalable Influence Maximization:\\ All Marginal Gain Estimates for the Cost of One}

\author{
Matic Požar\\
UP FAMNIT, University of Primorska\\
Koper, Slovenia\\
\texttt{matic.pozar@upr.si}
}

\date{}

\begin{document}

\maketitle


\begin{abstract}
Influence maximization (IM), the problem of selecting a small set of initially activated vertices to maximize the expected influence spread in a network, has applications in viral marketing and public-health interventions. Under the widely studied independent cascade (IC) model, IM is NP-hard, exact influence evaluation is \#P-hard, and Monte Carlo estimation is computationally expensive. While much of the research on scalable IM has focused on structural heuristics and reverse-reachability (RR) sampling, we demonstrate that analytical approximations of the diffusion process can provide highly competitive alternatives.

We develop a gradient-based framework that exploits differentiable analytical influence approximations to obtain estimates of all marginal influence gains simultaneously, reducing the computational cost from $O(n(n+m))$ to $O(n+m)$ for fixed propagation horizons, where $n$ and $m$ denote the number of vertices and edges, respectively. Combined with a proportional batching strategy, this yields influence maximization algorithms with an overall time complexity of $O((n+m)\log(K+1))$, where $K$ is the seed budget. We further exploit the modularity of the ALE influence approximation to eliminate iterative seed selection altogether.

We evaluate the resulting algorithms across 263 benchmark configurations, encompassing four synthetic network families and nine real-world networks with up to 2.39 million vertices and 30.62 million directed edges. Our IPL-based algorithm achieves a median influence shortfall of only $0.644\%$ relative to the best observed completed method, outperforming SKIM in both influence spread and runtime in 231 of 263 configurations. Our DMP-based algorithm achieves a median speedup of $20\times$ over completed IMM executions while generally maintaining comparable influence quality. Furthermore, ALE consistently exhibits subsecond selection runtimes and outperforms conventional structural heuristics and the tested OPIM-C-style and SUBSIM configurations in influence quality in a large majority of cases. These findings suggest that analytical approximations of the IC model deserve substantially greater attention in the development of scalable IM algorithms.
\end{abstract}

\section{Introduction}

Influence maximization (IM), formalized by \citet{im_kempe2003}, asks for a small set of initially activated vertices in a network so as to maximize the expected influence spread. Applications include viral marketing, recommendation systems, misinformation mitigation, and public-health interventions \citep{viralmarketing_richardson2002,linkrecommendation_coro2019,recommendation_sun2023,misinformation_budak2011,healthintervention_yadav2018}. The problem is defined with respect to a diffusion model specifying how influence propagates through the network. One of the most widely studied such models is the independent cascade (IC) model \citep{ic_goldenberg2001,im_kempe2003}, in which influence propagates along directed edges according to specified transmission probabilities.

Under the IC model, IM is NP-hard, but the expected influence spread is a monotone submodular function \citep{im_kempe2003}. Consequently, the classical greedy algorithm achieves a $(1-1/e)$-approximation by successively selecting vertices with the largest marginal influence gains. The principal computational obstacle is that evaluating influence spread exactly is itself \#P-hard \citep{mia_chen2010}, while estimating marginal gains through repeated Monte Carlo simulations is prohibitively expensive on large networks. Moreover, even if an efficient influence estimator is available, straightforward greedy selection requires its repeated evaluation for every candidate vertex at every selection step. The scalability of greedy IM therefore depends not only on the cost of estimating influence, but also on how efficiently these estimates can be exploited during optimization.

Two broad approaches to overcoming this obstacle have emerged. Sampling-based methods, most notably reverse-reachability methods, transform IM into a sampled coverage problem, enabling efficient seed selection while retaining probabilistic approximation guarantees. Analytical and heuristic methods instead exploit structural properties of the network or computationally inexpensive approximations of the diffusion process. Although the latter generally lack comparable approximation guarantees, they avoid the computational and memory costs associated with generating and maintaining large collections of samples. Their practical competitiveness ultimately depends on the tradeoff between the quality of the resulting seed sets and the computational resources required to obtain them.

In this work, we develop a gradient-based framework for highly scalable influence maximization that exploits analytical approximations of the IC model. The starting point is an observation previously exploited for exact influence maximization in \citet{exactgreedy_pozar2026}: the partial derivatives of the multilinear extension of the influence function, evaluated at the indicator vector of a deterministic seed set, coincide with the marginal influence gains of adding individual vertices. If the influence function is evaluated through a differentiable computational procedure, reverse-mode differentiation enables the simultaneous computation of all these derivatives at an asymptotic cost comparable to that of a single influence evaluation.

We extend this computational principle to analytical approximations of IC influence. Rather than evaluating an influence approximation separately for each candidate vertex, we differentiate it with respect to the initial activation probabilities, thereby obtaining estimates of all marginal gains simultaneously. For the finite-horizon analytical approximations considered in this work, this reduces the cost of scoring all candidates from $O(n(n+m))$ to $O(n+m)$, where $n$ and $m$ denote the number of vertices and edges, respectively, and the propagation horizon is treated as a fixed parameter. Unlike the derivatives of the exact multilinear extension, however, the derivatives of these approximations need not coincide with their discrete marginal gains. Their usefulness for influence maximization therefore depends on how effectively they identify promising candidate seeds.

We further reduce the computational cost of greedy selection through proportional batching. Instead of recomputing marginal-gain estimates after every selected vertex, we select progressively larger batches of vertices using a common set of estimates, with the maximum batch size proportional to the target seed-set size. Combining simultaneous marginal-gain estimation with this batching strategy reduces the overall time complexity from $O(Kn(n+m))$ for straightforward analytical greedy selection to $O((n+m)\log(K+1))$, where $K$ denotes the seed budget. This improvement comes at the cost of introducing additional approximation error, as the marginal-gain estimates are not updated to account for interactions between vertices selected within the same batch.

Our main contributions are as follows:

\begin{enumerate}
    \item \textbf{Simultaneous marginal-gain estimation.} We develop a general gradient-based framework for exploiting differentiable influence approximations in greedy IM. For the analytical approximations considered in this work, estimates of all candidate marginal gains can be obtained simultaneously in $O(n+m)$ time, reducing the complexity of sequential greedy selection from $O(Kn(n+m))$ to $O(K(n+m))$.

    \item \textbf{Proportional batching.} We introduce a proportional batching strategy that further reduces the complexity of gradient-based greedy selection to $O((n+m)\log(K+1))$.

    \item \textbf{Analytical IM algorithms.} We instantiate the proposed framework using the ALE, IPL, and DMP influence approximations. We exploit the modularity of the ALE objective to eliminate iterative seed selection altogether, obtaining an inexpensive IM heuristic. The IPL- and DMP-based algorithms account for interactions between selected seeds and offer progressively more computationally intensive alternatives. Across our benchmark, batched IPL achieves a median influence shortfall of only $0.644\%$ relative to the best observed completed method, while outperforming SKIM in both influence spread and runtime in 231 of 263 configurations. Batched DMP with a propagation horizon of five achieves a median speedup of $20\times$ over completed IMM executions.

    \item \textbf{Experimental evaluation.} We evaluate 19 algorithmic configurations across 263 graph--seed-budget configurations, comprising four synthetic network families and nine real-world networks, with up to 2.39 million vertices, 30.62 million directed edges, and seed budgets of up to 2,395. Our results demonstrate the competitiveness of analytical IM methods against established heuristic, analytical, and sampling-based algorithms, including IRIE, IMM, SKIM, OPIM-C-style sampling, and SUBSIM. We examine how their relative performance varies across network structures and transmission-probability models.
\end{enumerate}

\section{Preliminaries and Related Work}
\label{sec:preliminaries}

\subsection{Independent Cascade Model and Influence Maximization}
\label{subsec:ic_model_and_im}

Let $G=(V,E)$ be a simple directed graph, where $n=|V|$ and $m=|E|$. Each edge $e\in E$ has a transmission probability $p_e\in[0,1]$. Given a seed set $S\subseteq V$, the IC process starts with precisely the vertices in $S$ active at time $0$. At each subsequent time step, vertices activated in the preceding step independently attempt to activate each currently inactive out-neighbor, succeeding with the probability of the corresponding edge. Each newly activated vertex makes these attempts only once.

Equivalently, the IC process admits a \emph{live-edge representation}. A random graph $G'\sim G$ is obtained by retaining each edge $e$ independently with probability $p_e$. The vertices eventually activated from $S$ are exactly those reachable from $S$ in $G'$. Let $X_u^{G'}(S)$ indicate whether $u$ is reachable from $S$ in $G'$. The activation probability of $u$ and the expected influence spread are
\[
    p_u^{(\infty)}(S)
    =\mathbb{E}_{G'\sim G}[X_u^{G'}(S)],
    \qquad
    \sigma(S)=\sum_{u\in V}p_u^{(\infty)}(S).
\]
We also write $p_u^{(t)}(S)$ for the probability that $u$ is active by time $t$, and $\sigma_t(S)=\sum_{u\in V}p_u^{(t)}(S)$ for the corresponding expected spread.

For a seed budget $K$, the IM problem is
\[
    S^*\in\argmax_{\substack{S\subseteq V, ~|S|\le K}}\sigma(S).
\]
Its NP-hardness follows, for example, from a reduction from Set Cover \citep{im_kempe2003}. Exact evaluation of $\sigma(S)$ is also \#P-hard \citep{mia_chen2010}; in fact, the latter difficulty already arises after three time steps.

\begin{theorem} \label{thm:t=3_is_phard}
    Computing the expected influence spread in the IC model is \#P-hard even when the number of time steps is restricted to three.
\end{theorem}
\begin{proof}
    We reduce from counting independent sets in bipartite graphs, which is \#P-hard \citep{bipsharpp_provan1983}.

    Given a bipartite graph $(L,R,E)$, orient every edge from $L$ to $R$. Add a source $s$ with an edge to every vertex in $L$ and a sink $t$ with an incoming edge from every vertex in $R$. Assign transmission probability $1/2$ to the edges leaving $s$ and entering $t$, and probability $1$ to all other edges. Take $S=\{s\}$.

    Each realization of the probabilistic edges corresponds to a pair $(X,Y)$ with $X\subseteq L$ and $Y\subseteq R$: the vertices in $X$ have live incoming edges from $s$, and those in $Y$ have live outgoing edges to $t$. Every pair occurs with probability $2^{-|L|-|R|}$. The sink $t$ is activated within three steps if and only if there is an edge $(u,v)\in E$ with $u\in X$ and $v\in Y$. Thus, $t$ remains inactive exactly when $X\cup Y$ is an independent set of the original bipartite graph.

    Let $I$ be the number of such independent sets. Then
    \[
        p_t^{(3)}(\{s\})=1-\frac{I}{2^{|L|+|R|}}.
    \]
    The activation probabilities of all remaining vertices are easy to compute: $s$ is always active; each vertex of $L$ is active with probability $1/2$; and each $v\in R$ is active with probability $1-2^{-\deg(v)}$. Consequently,
    \[
        \sigma_3(\{s\})
        =1+\frac{|L|}{2}
        +\sum_{v\in R}\bigl(1-2^{-\deg(v)}\bigr)
        +1-\frac{I}{2^{|L|+|R|}}.
    \]
    All terms other than $I$ are computable in polynomial time, so an exact algorithm for $\sigma_3(\{s\})$ would also count independent sets in bipartite graphs. This proves the claim.
\end{proof}

In contrast, $\sigma_t(S)$ can be evaluated exactly in $O(n+m)$ arithmetic operations for $t\le 2$, using the independence of the relevant activation events. This is also reflected in the first two iterations of the dynamic message passing (DMP) equations \citep{dmp_lokhov2019}.

The influence function $\sigma$ is monotone and submodular \citep{im_kempe2003}. Writing
\[
    \Delta(v\mid S)=\sigma(S\cup\{v\})-\sigma(S)
\]
for the marginal gain of $v\notin S$, submodularity states that $\Delta(v\mid A)\ge\Delta(v\mid B)$ whenever $A\subseteq B$ and $v\notin B$. The classical greedy algorithm repeatedly adds a vertex of maximum marginal gain, yielding
\[
    \sigma(S_{\mathrm{greedy}})
    \ge \left(1-\frac{1}{e}\right)\sigma(S^*)
\]
when the gains are computed exactly. The same guarantee does not automatically apply when candidate scores are approximate.

\subsection{Influence Estimation}
\label{subsec:influence_estimation}

A standard estimator of IC influence spread is Monte Carlo (MC) simulation. Given $M$ independent live-edge realizations $G'_1,\ldots,G'_M$, define
\[
    \widehat\sigma_{\mathrm{MC}}(S)
    =\frac{1}{M}\sum_{i=1}^{M}
      \bigl|\operatorname{Reach}_{G'_i}(S)\bigr|,
\]
where $\operatorname{Reach}_{G'}(S)$ is the set reachable from $S$ in $G'$. This estimator is unbiased and consistent, but using it to evaluate many candidate sets across many greedy iterations is expensive.

Reverse-reachability sampling offers a more reusable form of stochastic influence estimation \citep{rr_borgs2014}. For a target vertex $u$ and a random live-edge realization, the \emph{reverse-reachable} (RR) set of $u$ contains every vertex that can reach $u$. A random RR set $RR$ is generated by selecting $u$ uniformly from $V$ and then constructing its RR set. The key identity is
\[
    \Pr(RR\cap S\neq\varnothing)=\frac{\sigma(S)}{n}.
\]
Hence, for $M$ independent RR sets $RR_1,\ldots,RR_M$,
\[
    \widehat\sigma_{\mathrm{RR}}(S)
    =\frac{n}{M}\sum_{i=1}^{M}
      \mathbf{1}[RR_i\cap S\neq\varnothing]
\]
is an unbiased estimator of $\sigma(S)$. The same RR samples can be reused to score candidate seeds, reducing IM to a sampled maximum-coverage problem.

Analytical estimators take a different route: instead of sampling diffusion outcomes, they approximate activation probabilities using local propagation structures or iterative equations. DMP, for example, updates messages along directed edges to approximate the evolution of IC activation probabilities \citep{dmp_lokhov2019}. Such methods can evaluate an approximate spread efficiently, but their accuracy depends on the network structure and the extent to which different propagation paths overlap. Their deterministic and often differentiable computations make them particularly suitable for the approach investigated here.

\subsection{Related Influence Maximization Algorithms}
\label{subsec:related_work}

\paragraph{Sampling-based greedy algorithms.}
The original greedy method of \citet{im_kempe2003} estimates candidate gains by repeated MC simulation. CELF \citep{celf_leskovec2007} and CELF++ \citep{celfpp_goyal2011} exploit submodularity to avoid reevaluating candidates whose previous gains remain valid upper bounds. NewGreedy \citep{newgreedy_chen2009} instead reuses sampled live-edge realizations to estimate the gains of multiple candidates more efficiently. These improvements reduce the cost of MC-based greedy selection without changing its basic selection rule.

\paragraph{Reverse-reachability and sketch-based algorithms.}
Building on reverse influence sampling \citep{rr_borgs2014}, TIM and TIM+ \citep{tim_tang2014} develop efficient RR-based IM algorithms with approximation guarantees. IMM \citep{imm_tang2015} uses martingale concentration bounds to control the number of RR samples needed, then greedily selects vertices covering the most previously uncovered RR sets. OPIM-C \citep{opimc_tang2018} uses progressive sampling together with an independent validation sample to assess solution quality and determine whether further sampling is necessary. SUBSIM \citep{subsim_guo2020,subsim_guo2022} targets the cost of generating RR sets themselves, using efficient subset sampling of live incoming neighbors and techniques to avoid unnecessary exploration.

SKIM \citep{skim_cohen2014} also exploits sampled reachability, but represents it through compact bottom-$k$ reachability sketches. It estimates residual influence from these sketches and updates them as seeds are selected, avoiding the need to store large collections of explicit reachability sets.

\paragraph{Analytical and structural heuristics.}
At the inexpensive end of the spectrum are degree-based and other centrality heuristics. Degree Discount and weighted variants account, approximately, for the overlap introduced by already selected seeds \citep{newgreedy_chen2009}. More elaborate analytical methods replace direct simulation with approximations of the diffusion process. MIA \citep{mia_chen2010}, for example, restricts attention to influential paths and local tree-like structures, while methods such as SPM, IPA, and EaSyIM use other simplified propagation calculations \citep{spm_kimura2006,IPA_kim2013,EaSyIM_galhotra2016}. IRIE \citep{irie_jung2012} is particularly relevant: it estimates influence rankings through a system of linear propagation equations and discounts vertices according to the influence already attributed to selected seeds. Unlike methods that estimate one candidate at a time, IRIE updates scores for all candidates jointly.

\paragraph{Differentiable influence optimization.}
Analytical diffusion models have also been used in continuous optimization problems. For example, \citet{dmp_control_lokhov2017} employ forward--backward DMP equations to optimize the deployment of spreading resources. Our approach builds on the differentiation of the exact IC influence function in \citet{exactgreedy_pozar2026}, extending it to analytical approximations for greedy seed selection.

\paragraph{Learning-based approaches.}
Learning-based IM methods constitute a further design direction, with examples including \citep{gcomb_manchanda2020,gnnim_panagopoulos2024,BiGDN_wenlong2025,topicaware_Halal2025,MaDGNN_Yang2026}. Their evaluation raises additional questions about generalization across networks and diffusion regimes, as well as whether training and preprocessing costs should be amortized over multiple IM queries. We therefore leave them outside the main benchmark. During preliminary work, we also investigated DeepIM \citep{deepim_ling2023}, but were unable to reproduce its reported performance and therefore excluded it from the benchmark.
\section{Gradient-Based Influence Maximization}
\label{sec:gradient_framework}

\subsection{Differentiable Approximations and Marginal Gains}
\label{subsec:approximation_functions}

We consider a continuous representation of the initial seed set. Let $x\in[0,1]^n$, where $x_v$ denotes the probability that vertex $v$ is initially activated, independently of all other vertices. Equivalently, we may introduce an auxiliary source vertex $s$, which is initially active and has an outgoing edge $(s,v)$ with transmission probability $x_v$ for every $v\in V$. The resulting random initial seed set is distributed according to the independent Bernoulli probabilities specified by $x$.

Let $F:[0,1]^n\rightarrow\mathbb{R}$ denote the corresponding exact expected influence spread, excluding the auxiliary source. For a deterministic seed set $S$, let $\mathbf{1}_S$ denote its indicator vector. Then
\[
F(\mathbf{1}_S)=\sigma(S).
\]

The function $F$ is the multilinear extension of $\sigma$. In particular, for every $v\notin S$, it satisfies
\[
\frac{\partial F}{\partial x_v}(\mathbf{1}_S)
=
F(\mathbf{1}_{S\cup\{v\}})
-
F(\mathbf{1}_S)
=
\Delta(v\mid S).
\]

As observed in \citet{exactgreedy_pozar2026}, this identity has an important computational consequence. If the influence function can be evaluated by a differentiable computational procedure, reverse-mode automatic differentiation allows all its partial derivatives to be computed simultaneously, with asymptotic computational cost comparable to a single evaluation of the underlying function. In contrast, straightforward evaluation of all marginal gains requires a separate influence computation for each candidate vertex.

In \citet{exactgreedy_pozar2026}, this observation was exploited in conjunction with an exact IC influence computation, yielding exact marginal gains. Here, we consider a more general setting in which the exact influence function is replaced by a computationally efficient approximation
$
\widehat F:[0,1]^n\rightarrow\mathbb{R}.
$
For a given seed set $S$, we use
\[
\boxed{
\widehat{\Delta}(v\mid S)
=
\frac{\partial\widehat F}{\partial x_v}
(\mathbf{1}_S)
}
\]
as an estimate of the true marginal gain $\Delta(v\mid S)$.

Unlike the exact influence function $F$, the approximation $\widehat F$ need not be multilinear. Consequently, its derivatives do not necessarily coincide with its own discrete marginal differences.
Thus, differentiation introduces an additional approximation beyond the error already present in $\widehat F$. Nevertheless, it provides an efficient mechanism for obtaining the scores of all candidate vertices simultaneously.

The effectiveness of this approach depends on the quality of the underlying influence approximation and, more specifically, on how well its derivatives preserve the relative ordering of the true marginal gains.

\subsection{Influence Approximation Methods}
\label{sec:approximations}

We consider three analytical approaches to approximating the IC influence function: the Aggregated Linear Effect (ALE) model \citep{ale_bota2013}, its modified variant with inter-layer overlap corrections, which we refer to as IPL \citep{icframework_pozar2026}, and Dynamic Message Passing (DMP) \citep{dmp_lokhov2019}.

All three methods approximate the propagation of influence through a finite number of iterations and can be evaluated in $O(T(n+m))$ time, where $T$ denotes the number of propagation iterations. Throughout this work, we treat $T$ as a fixed parameter, independent of the graph size. Consequently, a single evaluation of any of these approximations requires $O(n+m)$ time.

\subsection{Aggregated Linear Effect}
\label{subsec:ale}

The ALE model \citep{ale_bota2013} provides a particularly simple approximation of influence propagation by treating the contributions of different propagation paths additively.

Let $A\in\mathbb{R}^{n\times n}$ denote the weighted adjacency matrix, with $A_{vu}=p_{uv}$ if $(u,v)\in E$ and zero otherwise. The finite-horizon ALE approximation is defined by
\[
\widehat F_{\mathrm{ALE}}(x)
=
\mathbf{1}^{\mathsf T}
\left(
\sum_{t=0}^{T}A^t
\right)x.
\]
Thus, the expected influence is approximated by accumulating the contributions of all directed walks of length at most $T$, weighted by their transmission probabilities.

ALE counts weighted walks additively, without correcting for repeated activation or feedback along cycles.

Defining
\[
c=
\sum_{t=0}^{T}(A^{\mathsf T})^t\mathbf{1},
\quad\text{ we obtain }\quad
\widehat F_{\mathrm{ALE}}(x)
=
\sum_{v\in V}c_vx_v.
\]
Consequently, the ALE influence approximation is \emph{modular}: the contribution of each seed is independent of the other selected seeds.

It follows that IM under this approximation can be solved exactly by selecting the $K$ vertices with the largest coefficients $c_v$. These coefficients can be computed simultaneously through $T$ sparse matrix-vector multiplications, requiring $O(T(n+m))$ time.

\subsection{Independent Propagation Layer}
\label{subsec:ipl}

The Independent Propagation Layer (IPL) approximation is based on a modified ALE procedure that attempts to correct for repeated influence across different propagation steps \citep{icframework_pozar2026}.

Similarly to ALE, IPL propagates influence using additive contributions from incoming edges. However, rather than summing the influence generated at different steps, it maintains an estimate of the probability that each vertex has remained inactive.

Let $h^{(0)}=x$, and define the propagated influence at step $t\ge 1$ by
\[
h_v^{(t)}
=
\min\left\{
1,\,
\sum_{u:(u,v)\in E}
p_{uv}h_u^{(t-1)}
\right\}.
\]
The corresponding estimate of the activation probability of vertex $v$ is
\[
\widehat p_v^{(T)}
=
1-
\prod_{t=0}^{T}
\left(1-h_v^{(t)}\right),
\]
giving the influence approximation
\[
\widehat F_{\mathrm{IPL}}(x)
=
\sum_{v\in V}
\widehat p_v^{(T)}.
\]

The multiplicative survival term accounts approximately for the fact that influence reaching a vertex during different propagation steps should not be counted independently toward its final activation. Unlike DMP, IPL still aggregates incoming contributions additively within each propagation step.

The resulting influence approximation is no longer modular. The estimated contribution of a vertex depends on the other selected seeds, and selecting the $K$ vertices with the largest singleton influence estimates is therefore not generally optimal under the IPL objective.

The approximation is differentiable almost everywhere, and reverse-mode differentiation computes its gradient in $O(T(n+m))$ time. At the clipping threshold, we use the derivative convention of zero at and above one.

\subsection{Dynamic Message Passing}
\label{subsec:dmp}

Dynamic Message Passing (DMP) approximates IC propagation through a system of recursive probability equations \citep{dmp_lokhov2019}.

Unlike ALE and IPL, DMP maintains a separate propagation message for each directed edge. Let $\mu_{u\rightarrow v}^{(t)}$ denote the estimated probability that $u$ has been activated by iteration $t$ in a cavity process in which the contribution from $v$ back to $u$ is excluded.

Starting from
$
\mu_{u\rightarrow v}^{(0)}=x_u,
$
the messages are updated according to
\[
\mu_{u\rightarrow v}^{(t+1)}
=
1-(1-x_u)
\prod_{\substack{w:(w,u)\in E\\w\neq v}}
\left(
1-p_{wu}\mu_{w\rightarrow u}^{(t)}
\right).
\]

After $T$ message-passing iterations, the final vertex update approximates activation by time $T+1$:
\[
\widehat p_u
=
1-(1-x_u)
\prod_{w:(w,u)\in E}
\left(
1-p_{wu}\mu_{w\rightarrow u}^{(T)}
\right),
\]
and the corresponding influence approximation is
\[
\widehat F_{\mathrm{DMP}}(x)
=
\sum_{u\in V}\widehat p_u.
\]

Unlike IPL, DMP aggregates incoming activation probabilities multiplicatively and uses cavity messages to suppress immediate feedback along reciprocal edges.

DMP is exact on trees but does not eliminate dependencies arising from longer cycles and overlapping paths.

Using appropriate data structures, each message-passing iteration can be implemented in $O(n+m)$ time. However, maintaining edge-specific messages and differentiating through the recursive updates introduces greater computational and memory overhead than ALE or IPL.

\subsection{Seed Selection Algorithms}
\label{sec:seed_selection}

We distinguish between two selection strategies: sequential greedy selection, which recomputes the candidate scores after every selected seed, and batched greedy selection, which reuses the same scores to select multiple seeds before updating the approximation.

Throughout this section, let $\widehat F$ denote a differentiable influence approximation and write
\[
g_v(S)
=
\frac{\partial\widehat F}{\partial x_v}
(\mathbf{1}_S)
\]
for the corresponding estimated marginal gain of vertex $v\notin S$.
\paragraph{Sequential greedy selection.}

The most straightforward approach follows the classical greedy algorithm. Starting with $S=\varnothing$, we repeatedly select
\[
v^*
\in
\argmax_{v\in V\setminus S}g_v(S)
\]
and update $S\leftarrow S\cup\{v^*\}$ until $|S|=K$.

For the analytical methods considered here, computing the complete gradient requires $O(T(n+m))$ time. With a fixed propagation horizon $T$, sequential gradient-based greedy selection therefore runs in
$O(K(n+m))$
time.

This improves upon the straightforward implementation that separately evaluates the influence approximation for every candidate at every greedy iteration, requiring $O(Kn(n+m))$ time.
Unlike exact greedy IM, however, the selection rule uses approximate marginal gains and therefore does not automatically inherit the classical $(1-1/e)$ approximation guarantee.

\paragraph{Batched greedy selection.}

Although simultaneous differentiation eliminates the need for separate candidate evaluations, sequential greedy selection still requires $K$ complete gradient computations.

We therefore consider a batched variant that selects multiple vertices using the same gradient. Given the current seed set $S$, we compute $g_v(S)$ for all remaining candidates and select a batch $B\subseteq V\setminus S$ consisting of the vertices with the largest scores. The seed set is then updated to $S\cup B$, after which the gradient is recomputed.

The main disadvantage of batching is that candidate scores become increasingly outdated as vertices are added. In particular, selecting several vertices based on their individual marginal gains may overestimate their combined contribution because their influence regions can overlap.

To quantify the loss introduced by simultaneous seed selection, we consider the true IC influence function $\sigma$. For a seed set $S\subseteq V$ and a batch $B\subseteq V\setminus S$, define the \emph{batch overlap loss} as
\[
\mathcal E(S,B)
=
\sum_{v\in B}\Delta(v\mid S)
-
\bigl(\sigma(S\cup B)-\sigma(S)\bigr).
\]
By submodularity, $\mathcal E(S,B)\ge 0$. The following result establishes an additional property specific to the coverage structure of IC influence.

\begin{theorem}[Monotonicity of batch overlap loss]
\label{thm:batch_overlap}
Let $\sigma$ be the influence function of the independent cascade model. For any batch $B\subseteq V$ and seed sets
\[
S\subseteq S'\subseteq V\setminus B,\quad \text{we have}\quad 0\le \mathcal E(S',B)\le \mathcal E(S,B).
\]
In particular, the absolute overlap loss of any fixed batch is nonincreasing as the existing seed set grows.
\end{theorem}

\begin{proof}
Consider a fixed live-edge realization $G'$ of the IC model. For each vertex $v$, let $R_v$ denote the set of vertices reachable from $v$ in $G'$, and write
$R_S=\bigcup_{v\in S}R_v.$
For a fixed batch $B$, define
$r_B(u)=|\{v\in B:u\in R_v\}|,$
the number of vertices in $B$ that can reach $u$.

The marginal contribution of $v\in B$ relative to $S$ in this realization is
$|R_v\setminus R_S|,$
whereas the combined contribution of $B$ is
$\left|\bigcup_{v\in B}R_v\setminus R_S\right|.$
Consequently, the batch overlap loss in this realization equals
$$
\mathcal E_{G'}(S,B)
=
\sum_{v\in B}|R_v\setminus R_S|
-
\left|\bigcup_{v\in B}R_v\setminus R_S\right|
=
\sum_{u\notin R_S}\max\{r_B(u)-1,0\}.
$$

Since $S\subseteq S'$, we have $R_S\subseteq R_{S'}$. Moreover, $r_B(u)$ depends only on $B$ and the realization $G'$, and is therefore independent of $S$.

It follows that
$\mathcal E_{G'}(S',B)
\le
\mathcal E_{G'}(S,B).$
Taking expectations over all live-edge realizations yields
$\mathcal E(S',B)\le\mathcal E(S,B).$
Nonnegativity follows directly from the same representation.
\end{proof}

Theorem~\ref{thm:batch_overlap} motivates increasing batch sizes as selection progresses, since previously influenced vertices no longer contribute to the absolute overlap loss of a fixed batch. However, it does not establish that relative overlap loss decreases or that stale candidate rankings become more accurate. The batches selected by the algorithm also change between iterations. Increasing batch sizes is therefore a heuristic rather than a strategy with guaranteed improvement in approximation quality.

\paragraph{Proportional batching.}

We employ a geometric schedule in which the batch size initially equals two and doubles after each completed batch, up to a cap proportional to the seed-set budget.

Let $c\ge 1$ be a fixed parameter controlling the maximum batch size. At batching iteration $i=0,1,\ldots$, define
\[
b_i
=
\min\left\{
2^{i+1},
\max\left\{2,\left\lceil\frac{K}{c}\right\rceil\right\},
K-|S|
\right\}.
\]
The procedure is summarized in Algorithm~\ref{alg:proportional_batching}.

\begin{algorithm}[t]
\caption{Proportional batched greedy influence maximization}
\label{alg:proportional_batching}
\begin{algorithmic}[1]
\Require Graph $G$, budget $K$, differentiable influence approximation $\widehat F$, batch parameter $c$
\Ensure Seed set $S$ of size $K$
\State $S\gets\varnothing$
\State $i\gets 0$
\While{$|S|<K$}
    \State $g\gets\nabla\widehat F(\mathbf{1}_S)$
    \State $b\gets\min\left\{2^{i+1},
    \max\{2,\lceil K/c\rceil\},K-|S|\right\}$
    \State $B\gets$ the $b$ vertices in $V\setminus S$ with largest scores $g_v$
    \State $S\gets S\cup B$
    \State $i\gets i+1$
\EndWhile
\State \Return $S$
\end{algorithmic}
\end{algorithm}

Before reaching the batch-size cap, the batch sizes grow geometrically, requiring $O(\log(K+1))$ iterations. Once the cap $\lceil K/c\rceil$ is reached, at most $O(c)$ additional batches are needed.

Consequently, the total number of gradient computations is
$O(\log(K+1)+c).$
For fixed $c$, this reduces to $O(\log(K+1))$.
Since each gradient computation requires $O(T(n+m))$ time, the total cost of gradient evaluations is
$O\bigl(T(n+m)\log(K+1)\bigr).$
Candidate selection can be implemented using a priority structure that is rebuilt after each gradient computation and supports extracting a maximum-scoring candidate in $O(\log n)$ time. This contributes an additional
$O\bigl(n\log(K+1)+K\log n\bigr)$
time. For a fixed propagation horizon and fixed batching parameter $c$, the resulting overall complexity is therefore
\[
O\bigl((n+m)\log(K+1)+K\log n\bigr)
=
\boxed{
O\bigl((n+m)\log(K+1)\bigr),
}
\]
where the final simplification uses $K\le n$.

\section{Experimental setup}
\label{sec:experimental_setup}

We evaluate the proposed methods under the independent cascade (IC) model on synthetic and real-world networks. The benchmark comprises 263 graph--seed-budget configurations and 19 algorithmic configurations. Of the 14,991 method--repetition slots, 14,907 yielded influence evaluations; incomplete executions retain their recorded statuses.

\paragraph{Networks.}
We use four synthetic graph families: Erd\H{o}s--R\'enyi (ER), Barab\'asi--Albert (BA), Watts--Strogatz (WS), and stochastic block models (SBM). ER graphs contain exactly $8n$ distinct directed arcs sampled uniformly without replacement, excluding self-loops. BA graphs start from a five-vertex clique, with each new vertex attaching to four existing vertices by preferential attachment. WS graphs start from an undirected ring lattice of degree eight and rewire each forward lattice edge independently with probability $0.15$. BA and WS edges are represented by reciprocal directed arcs. SBM graphs have four approximately equal communities, with within-community connection probabilities approximately ten times the between-community probabilities; directed block-pair edge counts are calibrated to an expected out-degree of approximately eight. Self-loops and duplicates are removed. All families thus have average (out-)degree close to eight.

We consider $n\in\{10^4,3\cdot10^4,10^5,3\cdot10^5,10^6\}$, with three independent realizations per family at the first three sizes and two at the last two. Under heterogeneous activation probabilities, the seed budgets are
\[
\begin{array}{c|l}
n & K\\ \hline
10^4 & 10,20\\
3\cdot10^4 & 20,30,50,100,200,300\\
10^5 & 20,50,100,200,1000\\
3\cdot10^5 & 20,300\\
10^6 & 20,1000
\end{array}
\]
for every family and realization. Additional weighted-cascade experiments use $n=10^5$, three realizations per family, and $K\in\{20,200\}$, giving 64 synthetic graph--probability configurations overall.

We also use nine real-world networks from SNAP (Table~\ref{tab:benchmark_datasets}). Original edge directions are preserved in directed datasets; undirected edges are converted into reciprocal arcs. Self-loops and duplicate arcs are removed, and vertices are reindexed consecutively. Each dataset contributes one topology, shared between diffusion regimes where both are tested.

\begin{table*}[t]
\centering
\caption{Real-world benchmark datasets. The edge count $m$ refers to directed arcs after preprocessing; reciprocal representation stores each undirected edge as two arcs. The listed seed budgets apply to the heterogeneous model. WC is also evaluated on Pokec, WikiTalk, and YouTube with $K\in\{20,500\}$.}
\label{tab:benchmark_datasets}
\small
\begin{tabular}{lrrll}
\hline
Dataset & $n$ & $m$ & Representation & Heterogeneous-model seed budgets\\
\hline
Pokec       & 1,632,803 & 30,622,564 & Directed
            & $20,50,100,200,500,1000,1633$\\
WikiTalk    & 2,394,385 & 5,021,410 & Directed
            & $20,50,100,200,500,1000,2395$\\
YouTube     & 1,134,890 & 5,975,248 & Reciprocal
            & $20,50,100,200,500,1000,1135$\\
Epinions    & 75,879 & 508,837 & Directed
            & $20,50,100,200$\\
Enron       & 36,692 & 367,662 & Reciprocal
            & $20,50,100,200$\\
WikiVote    & 7,115 & 103,689 & Directed
            & $20,50,100,200$\\
Slashdot    & 82,168 & 870,161 & Directed
            & $20,50,100,200$\\
CondMat     & 23,133 & 186,878 & Reciprocal
            & $20,50,100,200$\\
citHepTh    & 27,769 & 352,768 & Directed
            & $20,50,100,200$\\
\hline
\end{tabular}
\end{table*}

\paragraph{Activation probabilities.}
We consider two probability regimes. In the heterogeneous model, each vertex $u$ independently receives a source-activity parameter $a_u\in\{0.2,1.8\}$, with equal probability, shared across its outgoing arcs. For each arc $(u,v)$, we set
\[
p_{uv}=1-\exp\!\left(-\frac{\lambda a_u}{d^-(v)}\right),
\qquad \lambda=1.2,
\]
where $d^-(v)$ is the destination's in-degree. The latent source activity introduces variation in transmission strength independently of topology, making degree alone a less reliable indicator of influence. Separately, the exponential form provides smooth control of diffusion strength through $\lambda$ while keeping transmission probabilities below one for finite $\lambda$. As a standard degree-normalized reference, we also consider weighted cascade (WC), where $p_{uv}=1/d^-(v)$. IC transmission events remain independent across directed arcs conditional on their assigned probabilities, including for reciprocal arcs. Topology and activity seeds are fixed independently of $\lambda$.

\paragraph{Compared algorithms.}
The 19 configurations cover analytical approximations, gradient-based greedy methods, heuristics, and sampling-based methods. The main figures focus on ALE (three-step approximation), DMP with $T\in\{2,5\}$, batched IPL with $T=3$, IMM ($\varepsilon=0.1$), SKIM (64/64), and IRIE (MIOA-based correction). DMP and batched IPL use proportional batching: starting with two seeds, the batch size doubles after each gradient computation up to $\max\{2,\lceil K/8\rceil\}$. The remaining configurations include sequential and overlap-corrected DMP/IPL variants, weighted degree and its discounted variant, one-round coverage greedy, random selection, an OPIM-C-style RR baseline, and SUBSIM. The implementations and results of all configurations, including the overlap-corrected variants not analyzed here, are available in the accompanying repository.

\paragraph{Evaluation and resource limits.}
Each algorithm is run three times per graph and seed budget, and its selected seed sets are evaluated using 10,000 Monte Carlo simulations of eventual IC spread, irrespective of the horizon used during selection. Common random numbers are used across seed sets. For the main comparisons, we report mean influence and median seed-selection runtime over three completed repetitions, with Monte Carlo uncertainty and completion counts where relevant. The supplementary real-world DMP runs at $T=1,3,4$ use a single selection execution per configuration, evaluated with the same 10,000-trial protocol.

Each selection attempt is limited to 1,800 seconds and 6 GiB of memory; RR-based methods additionally have caps of $8\cdot10^6$ RR sets and $8\cdot10^7$ total RR memberships. A failure or cap does not prevent other algorithms from running. For large budgets, IRIE may also be skipped when checkpoint-based extrapolation predicts infeasible runtime. Valid seed sets returned by capped methods can still be evaluated, but such runs are identified as incomplete and excluded from success-only aggregates; in particular, early termination may invalidate theoretical approximation guarantees.

\paragraph{Implementation and reproducibility.}
The OPIM-C-style and SUBSIM baselines are compiled from the SUBSIM authors' GitHub C++ implementation, using vanilla RR sampling and SUBSIM sampling, respectively; the former is not an independently sourced OPIM-C implementation. Both use $\varepsilon=0.1$, $\delta=1/n$, and no HIST acceleration. The remaining methods are implemented in Python with Numba-compiled numerical routines. The main runtime comparisons use a separate timing campaign on the same local workstation. The batching ablation uses historical selection timings, comparing only configurations with matching execution-host classifications. Source code, benchmark configurations, experimental measurements, and scripts for regenerating the figures and tables are available at \url{https://github.com/Matic054/Scalable-Influence-Maximization}.

\paragraph{Use of AI-assisted tools.}

OpenAI ChatGPT was used throughout the research process, primarily to assist with algorithm implementation, experimental methodology, literature reviews, scientific discussion, and manuscript preparation.

The principal scientific ideas originate from the author, with notable AI-assisted contributions including the proof of Theorem~\ref{thm:t=3_is_phard}, the formalization of Theorem~\ref{thm:batch_overlap}, the introduction of a proportional upper bound on batch size, the observation that the ALE objective is modular, and the exponential formulation of the edge-probability model used in the experimental evaluation.

The author takes full responsibility for the scientific contributions, mathematical arguments, experimental methodology, reported results, and final manuscript.

\section{Experimental Results}
\label{sec:results}

We separately examine inexpensive and more computationally intensive IM methods. The main figures focus on the three largest real-world networks: Pokec, WikiTalk, and YouTube.

\subsection{Highly Scalable Influence Maximization Methods}
\label{subsec:highly-scalable}

We compare ALE ($T=3$) with four inexpensive baselines: \emph{weighted degree}, which ranks vertices by the sum of their outgoing transmission probabilities; \emph{weighted degree discount}, which multiplies $1+\sum_{(u,v)\in E}p_{uv}$ by the probability that $u$ has not been directly activated by already selected seeds; \emph{one-round coverage}, which greedily optimizes exact one-step IC influence while accounting for overlap; and \emph{random selection}, which chooses a uniformly random $K$-vertex set. We also include OPIM-C-style and SUBSIM RR methods in this category because in many instances they are orders of magnitude faster than more computationally intensive influence-maximization methods, yet their seed quality is often no better than that of the simple heuristics.

\begin{table}[h]
\centering
\caption{Number of benchmark configurations in which ALE achieves greater influence spread, lower seed-selection runtime, or both, relative to each competitor. Only configurations with completed results for both methods are compared.}
\label{tab:ale_baseline_comparison}
\begin{tabular}{lccc}
\hline
Competitor & Influence & Runtime & Both \\
\hline
Weighted degree          & 212/263 & 230/263 & 184/263 \\
Weighted degree discount & 209/263 & 120/263 & 82/263  \\
One-round coverage       & 209/263 & 194/263 & 141/263 \\
Random selection         & 263/263 & 0/263   & 0/263   \\
OPIM-C-style             & 203/257 & 257/257 & 203/257 \\
SUBSIM                   & 200/257 & 255/257 & 199/257 \\
\hline
\end{tabular}
\end{table}

Table~\ref{tab:ale_baseline_comparison} shows that ALE produces higher influence than each non-random heuristic in approximately 80\% of configurations. Weighted degree discount is faster than ALE in a majority of cases (143/263), but these differences are generally of little practical consequence at the observed time scale: ALE's median selection time is $0.0089$~s, and even its maximum across all 263 configurations is only $0.703$~s. Given its consistently subsecond runtime and generally superior influence quality, we regard ALE as the strongest overall choice among the inexpensive methods considered.

\begin{figure*}[t]
\centering
\includegraphics[width=\textwidth]{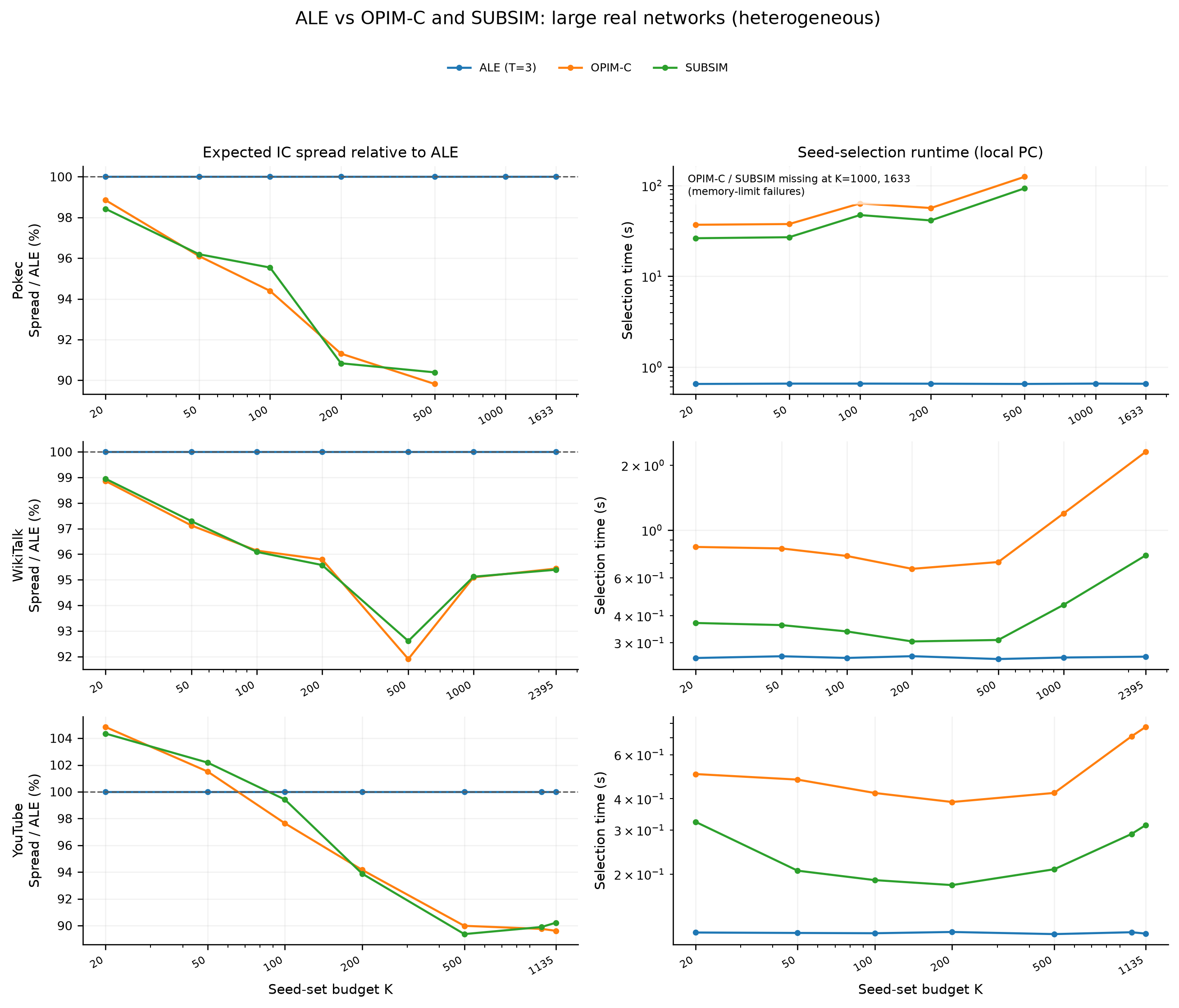}
\caption{ALE, OPIM-C-style RR sampling, and SUBSIM on Pokec, WikiTalk, and YouTube with heterogeneous activation probabilities. Left: influence spread relative to ALE (100\%). Right: seed-selection runtime in seconds, on a logarithmic vertical axis. Missing Pokec points correspond to incomplete RR-method executions.}
\label{fig:ale_rr_large_real_heterogeneous}
\end{figure*}

ALE also compares favorably with the two RR-based methods: it attains greater influence in 203 of 257 complete comparisons against OPIM-C-style sampling and 200 of 257 against SUBSIM, while being faster in 257 and 255 comparisons, respectively. Figure~\ref{fig:ale_rr_large_real_heterogeneous} illustrates the tradeoff on the three largest real-world networks under heterogeneous probabilities. On Pokec and WikiTalk, ALE achieves higher influence at every comparable seed budget; on YouTube, OPIM-C-style sampling and SUBSIM have a modest advantage at the two smallest budgets, but ALE overtakes them for larger $K$. ALE's runtime remains nearly independent of $K$, whereas the RR methods generally become more expensive as the seed budget grows.

The overall quality comparison remains favorable to ALE under weighted cascade (WC), but runtime behavior can change markedly with the edge-probability model. On Pokec, at the common budgets $K=20$ and $K=500$, OPIM-C-style sampling is approximately $43$ and $93$ times slower under heterogeneous probabilities than under WC, respectively; the corresponding factors for SUBSIM are $40$ and $50$. The heterogeneous model retains bounded total incoming transmission mass, since $\sum_{u:(u,v)\in E}p_{uv}\leq 1.2\cdot1.8=2.16$ for every vertex $v$. The runtime differences demonstrate sensitivity to the transmission-probability model, although our comparison does not isolate heterogeneity from differences in overall diffusion strength. This sensitivity is particularly relevant to practical applications, where transmission probabilities need not conform to the simplified WC model. ALE still has a substantial speed advantage on heterogeneous Pokec, although the gap under WC is much smaller; on WC WikiTalk, SUBSIM is modestly faster than ALE.

The broad trend also holds on the other synthetic families and smaller real-world networks, with two notable exceptions. On heterogeneous Watts--Strogatz (WS) graphs, OPIM-C-style sampling and SUBSIM outperform ALE in 45 of 47 configurations, with median influence advantages of $6.53\%$ and $7.24\%$, respectively. Interestingly, under WC, ALE outperforms both in all six WS configurations. On the real-world cit-HepTh citation network, again under the heterogeneous model, OPIM-C-style sampling and SUBSIM outperform ALE at all four seed budgets, with median advantages of $12.60\%$ and $11.51\%$, respectively.

The comparison uses fixed, a priori parameters for both RR methods, rather than tuning their sampling budgets to match ALE's observed influence.

\subsection{Higher-Accuracy Influence Maximization Methods}
\label{subsec:high-inf}

We now consider methods that devote more computation to seed quality: IPL with horizon $T=3$, DMP with horizons $T\in\{1,2,3,4,5\}$, IRIE, IMM ($\varepsilon=0.1$), and SKIM (64/64). The IPL and DMP variants use the proportional batching scheme described in Section~\ref{sec:experimental_setup}. We also retain ALE ($T=3$) in the figures as a reference point from the highly scalable methods.

\begin{figure*}[t]
    \centering
    \includegraphics[width=\textwidth]{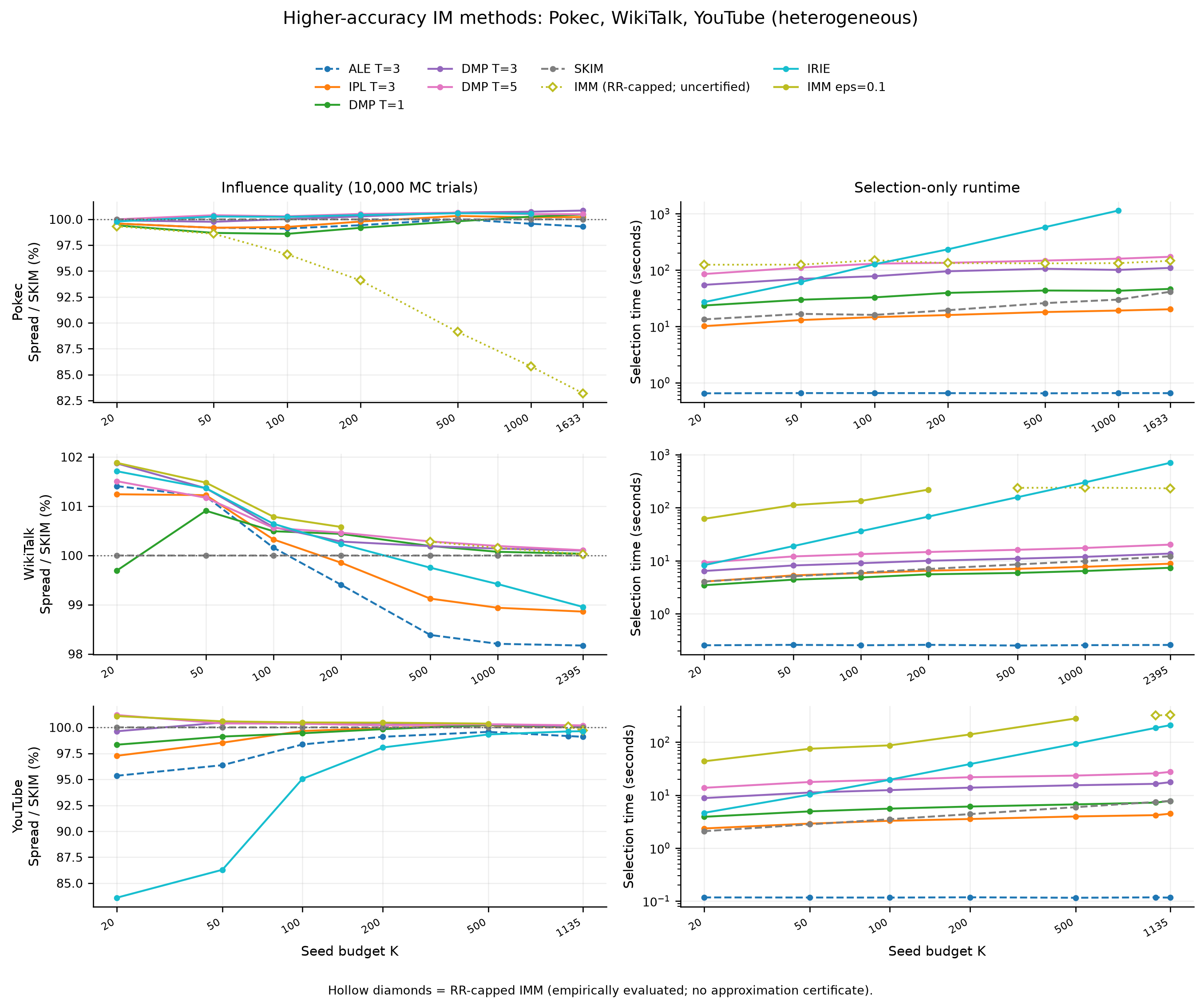}
    \caption{Influence spread relative to SKIM (left) and seed-selection runtime on a logarithmic scale (right) for the three largest real-world networks under heterogeneous activation probabilities. Influence is evaluated using $10{,}000$ Monte Carlo trials. Solid yellow markers indicate completed IMM ($\varepsilon=0.1$) executions; hollow yellow diamonds and dotted lines indicate empirically evaluated seed sets returned by IMM after reaching an RR resource cap. These capped executions are not certified and their runtimes are times to the cap, not times to successful completion.}
    \label{fig:accurate_large_real_heterogeneous}
\end{figure*}

Figure~\ref{fig:accurate_large_real_heterogeneous} illustrates the principal quality--runtime tradeoffs on Pokec, WikiTalk, and YouTube. A qualification concerning IMM is essential: although it was attempted on all 263 configurations, it completed without reaching the imposed RR-set or RR-membership limits on only 125. In the remaining 138 cases, an intermediate seed set was available and its spread could be evaluated empirically, but the prescribed sampling procedure had not finished and the corresponding approximation certificate was unavailable. These capped results are shown separately and excluded from completed-method comparisons. Throughout the higher-accuracy comparison, the best observed influence is the maximum measured spread among completed ALE, IPL, DMP, IRIE, IMM and SKIM configurations, including the additional DMP horizons where evaluated.

A particularly strong result is the tradeoff achieved by IPL. Across all 263 configurations, its median influence shortfall from the best observed completed method is only $0.644\%$ (mean $1.171\%$). Moreover, IPL outperforms SKIM in influence in 235/263 configurations and in runtime in 255/263, achieving both improvements in 231/263. All eight configurations in which SKIM is faster involve networks with at least one million vertices and small seed budgets ($K\in\{20,50\}$); the runtime difference exceeds a factor of $1.22$ in only one of these cases. IPL is also faster than IRIE in all 262 comparable completed configurations and faster than IMM in all 125 configurations in which IMM completes.

DMP provides an explicit mechanism for spending additional computation on more extensive propagation approximations. We evaluated $T=2$ and $T=5$ on the full set of 263 configurations, while the additional horizons $T=1,3,4$ were evaluated on the 51 real-world configurations using single selection runs rather than three-repetition timing medians.

\begin{figure*}[t]
    \centering
    \includegraphics[width=\textwidth]{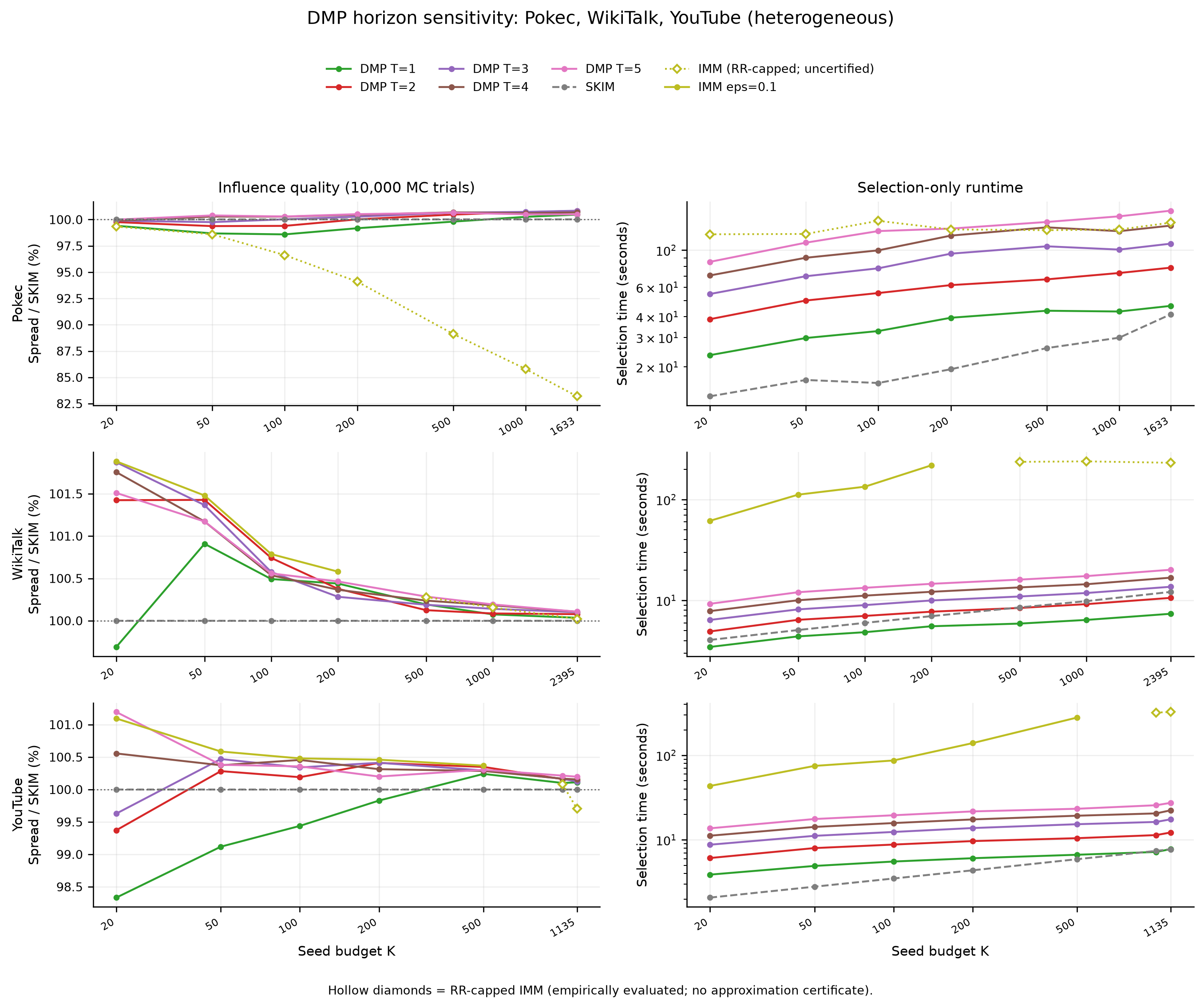}
    \caption{Sensitivity to DMP propagation horizon $T=1,\ldots,5$ on Pokec, WikiTalk, and YouTube under heterogeneous probabilities. Left: influence relative to SKIM. Right: seed-selection runtime (logarithmic scale). SKIM provides a common reference; IMM is included for comparison, with hollow diamonds denoting capped, uncertified runs as in Figure~\ref{fig:accurate_large_real_heterogeneous}.}
    \label{fig:dmp_horizons_large_real_heterogeneous}
\end{figure*}

Figure~\ref{fig:dmp_horizons_large_real_heterogeneous} shows that higher horizons can improve influence, but the gain is usually much smaller than the increase in runtime. Table~\ref{tab:dmp_horizon_discussion} summarizes this tradeoff on the 51 real-world configurations. The gains beyond $T=3$ are small relative to the additional runtime.

\begin{table}[h]
\centering
\caption{Influence shortfall from the best observed completed method
in the accurate-methods comparison and runtime relative to IPL
($T=3$), aggregated over the same 51 real-world configurations.
Each entry is a median or mean of within-configuration quantities.}
\label{tab:dmp_horizon_discussion}
\small
\setlength{\tabcolsep}{4pt}
\begin{tabular}{lrrr}
\hline
Horizon & Median gap & Mean gap & Time / IPL \\
\hline
$T=1$ & $0.377\%$ & $1.467\%$ & $2.19\times$ \\
$T=2$ & $0.173\%$ & $0.736\%$ & $2.74\times$ \\
$T=3$ & $0.115\%$ & $0.510\%$ & $3.89\times$ \\
$T=4$ & $0.109\%$ & $0.429\%$ & $4.94\times$ \\
$T=5$ & $0.112\%$ & $0.389\%$ & $6.15\times$ \\
\hline
\end{tabular}
\end{table}

On the full benchmark excluding Watts--Strogatz (WS) graphs, the median deviation from the best observed completed method is $0.323\%$ for DMP $T=2$ and $0.000\%$ for DMP $T=5$ (210 configurations). DMP $T=5$ attains the highest measured influence (including ties) in 116 of these 210 configurations.

On the 125 configurations in which IMM completes, DMP $T=5$ is faster in 123, with a $20\times$ median speedup. Excluding heterogeneous WS instances, the median relative influence difference (relative to IMM) is $+0.030\%$ in favor of DMP, with a median absolute relative difference of only $0.089\%$ across the remaining 101 configurations.

The qualitative picture carries over from real-world to synthetic networks, but there are two notable exceptions. First, under heterogeneous probabilities, DMP $T=5$ performs poorly on WS: across the 47 such configurations its median influence is only about $85.3\%$ of DMP $T=2$'s. Under weighted cascade, by contrast, the two horizons achieve essentially identical influence on WS. Second, SKIM (in the tested 64/64 configuration) is substantially less competitive in influence on some synthetic families under weighted cascade. For instance, at $n=100{,}000$ IPL achieves approximately $23\%$ and $24\%$ more influence than SKIM on Erd\H{o}s--R\'enyi and stochastic block model graphs, respectively.

Table~\ref{tab:batching_ablation} illustrates the tradeoff between computational efficiency and influence quality introduced by proportional batching. As expected, the computational advantage increases substantially with the seed budget, reaching median speedups of approximately $74$ and $78$ for IPL and DMP, respectively, at $K=1000$. Despite these substantial savings, the median influence loss remains below $0.3\%$ for both methods across the considered seed budgets. However, the mean and 95th percentile losses reveal that DMP is considerably more sensitive to batching, particularly on heterogeneous Watts--Strogatz networks. Across all comparable seed budgets, excluding heterogeneous WS, the mean influence loss introduced by batching is below $0.1\%$ for IPL $T=3$ and DMP $T\in\{2,5\}$.

\begin{table}[h]
\centering
\caption{Effect of proportional batching relative to sequential
gradient-based greedy selection for representative seed budgets.
Speedups and relative influence losses are computed within each
benchmark configuration and aggregated over configurations with
completed executions of both variants and matching execution-host
classifications.}
\label{tab:batching_ablation}
\small
\setlength{\tabcolsep}{4pt}
\begin{tabular}{lrrrrrr}
\hline
& \multicolumn{3}{c}{IPL ($T=3$)}
& \multicolumn{3}{c}{DMP ($T=5$)} \\
\cline{2-4}\cline{5-7}
Seed budget $K$ & 20 & 100 & 1000 & 20 & 100 & 1000 \\
\hline
Comparable cases & 76 & 33 & 22 & 76 & 32 & 17 \\
Median speedup
    & $2.79\times$ & $9.67\times$ & $74.31\times$
    & $2.79\times$ & $9.81\times$ & $78.05\times$ \\
Median loss (\%)
    & 0.000 & 0.000 & 0.052
    & 0.000 & 0.072 & 0.280 \\
Mean loss (\%)
    & 0.209 & 0.252 & 0.722
    & 2.389 & 2.457 & 3.540 \\
95th percentile loss (\%)
    & 0.730 & 1.800 & 3.819
    & 15.116 & 13.285 & 15.730 \\
\hline
\end{tabular}
\end{table}

\paragraph{Empirical Pareto frontier.}
Table~\ref{tab:pareto_discussion} reports how often each method is
nondominated in influence spread and seed-selection runtime among
all completed methods included in the main comparisons for the
same graph and seed budget.

\begin{table*}[t]
\centering
\caption{Number of configurations on the observed influence--runtime
Pareto frontier, among all main-comparison methods with completed
results. The denominator is the number of configurations in which
the method is available.}
\label{tab:pareto_discussion}
\begin{tabular}{lr@{\hspace{1.3em}}lr}
\hline
Method & Frontier / cases & Method & Frontier / cases \\
\hline
Random & $263/263$ & ALE & $244/263$ \\
Weighted degree & $20/263$ & Weighted degree discount & $164/263$ \\
One-round coverage & $54/263$ & OPIM-C-style & $10/257$ \\
SUBSIM & $10/257$ & IPL $T=3$ & $172/263$ \\
DMP $T=1$ & $24/51$ & DMP $T=2$ & $172/263$ \\
DMP $T=3$ & $27/51$ & DMP $T=4$ & $20/51$ \\
DMP $T=5$ & $127/263$ & IRIE & $86/262$ \\
IMM & $62/125$ & SKIM & $16/263$ \\
\hline
\end{tabular}
\end{table*}

The frontier counts are not an overall quality ranking. Random selection is nondominated because of its negligible selection time, despite producing poor seed sets. Moreover, minute differences in measured spread can determine whether a method is classified as dominated.

\section{Discussion}
\label{sec:discussion}

Our findings make a case for treating ALE as a standard baseline for fast influence maximization, at least on networks and diffusion regimes comparable to those examined here. Unlike RR methods run to their prescribed stopping criteria, ALE does not provide approximation guarantees for the IC objective merely because its surrogate can be optimized exactly. Characterizing the graph and probability classes for which its modular ranking is reliable is an open theoretical question.

The performance of DMP depends on both the propagation horizon and the selection schedule. DMP $T=5$ achieves a mean shortfall of only $0.148\%$ after excluding Watts--Strogatz (WS) networks, compared with $3.30\%$ on the full benchmark. The large batching losses on heterogeneous WS suggest that stale marginal estimates contribute to the deterioration, but do not establish that batching accounts for the entire difference between the two horizons. Adaptive selection of the propagation horizon and batch schedule may offer a better compromise than fixing either parameter in advance.

The diffusion model deserves more attention in experimental IM research. Widely used settings such as weighted cascade and small constant edge probabilities represent only part of the possible range of transmission strengths. When influence is weak, RR sets tend to be small, although a low optimal spread can increase the \emph{number} of RR samples needed for a prescribed relative accuracy. At stronger diffusion levels, individual RR sets may become much larger, while analytical approximations can be challenged by longer cascades and overlapping propagation paths. Our weighted-cascade and heterogeneous experiments illustrate the sensitivity of algorithm performance to the probability model. Broader probability distributions, including empirically estimated transmission probabilities, would help establish where different approximation and sampling techniques are most effective. Quality-matched tuning of sampling parameters would also clarify the runtime differences observed here.

\section{Conclusion}
\label{sec:conclusion}

We developed a gradient-based framework for influence maximization in which estimates of all marginal gains are computed simultaneously by differentiating analytical influence approximations. Combined with proportional batching, this yields seed-selection algorithms with time complexity $O((n+m)\log(K+1))$ for fixed propagation horizons and batching parameters.

Our experimental results establish ALE as a competitive inexpensive heuristic and demonstrate that IPL and DMP can achieve influence spread comparable to established analytical and sampling-based methods at substantially reduced computational cost. The observed exceptions, particularly on heterogeneous Watts--Strogatz networks, demonstrate that increasing the propagation horizon does not necessarily improve seed quality.

Our evaluation is limited by fixed algorithm parameters, synthetic transmission probabilities, resource-constrained executions, and Monte Carlo estimation uncertainty. Future work should investigate theoretical guarantees for gradient-based selection and adaptive strategies for choosing propagation horizons and batch sizes.


\bibliographystyle{plainnat}
\bibliography{references}

\end{document}